%% file: main.tex
\documentclass{article}
\PassOptionsToPackage{numbers,compress}{natbib}
\usepackage{iclr2026_conference,times}
\iclrfinalcopy
\input{config}

\title{Cross-Scenario Unified Modeling of\\User Interests at Billion Scale}

\author{%
    \bf Manjie Xu$^{\,1,3,*}$, Xin Jia$^{\,3,*}$, Cheng Chen$^{\,3,*}$, Jingyi Zhou$^{\,2,3}$, Chi Zhang$^{\,1,\,\textrm{\Letter}}$, Yongji Wu$^{\,3}$,\\
    \bf Zejian Wang$^{\,3}$, Kai Zuo$^{\,3,\dagger}$, Yibo Chen$^{\,3,\dagger}$, Xu Tang$^{\,3,\dagger}$, Yao Hu$^{\,3,\,\textrm{\Letter}}$, Yixin Zhu$^{\,1,\,\textrm{\Letter}}$ \\
    $^*$ equal contribution \, $^\dagger$ project lead \, $^\textrm{\Letter}$ corresponding author \\
    $^1$ Peking University \,
    $^2$ Fudan University \,
    $^3$ Xiaohongshu Inc.
}

\begin{document}
\maketitle

\begin{abstract}
User interests on \ac{ugc} platforms are inherently diverse, manifesting through complex behavioral patterns across heterogeneous scenarios such as search, feed browsing, and content discovery.
Traditional recommendation systems operate in isolated scenarios, optimizing business metrics within narrow contexts while neglecting valuable cross-scenario behavioral signals. This fragmented approach struggles to integrate advanced techniques like \acsp{llm} at billion-scale deployments, ultimately limiting the ability to capture holistic user interests across platform touchpoints.
We introduce \acs{method}, an \acs{llm}-enhanced hierarchical \underline{\textit{R}}ecommender \underline{\textit{E}}ngine for \underline{\textit{D}}iversified scenarios, tailored for industry-level \ac{ugc} recommendation systems. 
\acs{method} unifies user interest representations by aggregating and synthesizing actions from multiple behavioral contexts, enabling comprehensive item and user modeling. The framework features an \acs{llm}-powered architecture that delivers nuanced, multifaceted representations while maintaining deployment efficiency. A novel scenario-aware dense mixing and querying policy effectively fuses diverse behavioral signals to capture cross-scenario user intent patterns and express fine-grained, context-specific preferences during serving.
We validate \acs{method} through online A/B testing on hundreds of millions of users in \platform, demonstrating substantial performance gains in content recommendation and advertisement targeting tasks. We also introduce a million-scale sequential recommendation dataset, \acs{dataset}, for offline evaluation.
Our work advances unified user modeling, unlocking deeper personalization and fostering more meaningful user engagement in large-scale \ac{ugc} platforms.
\end{abstract}

\section{Introduction}

Modern \acf{ugc} platforms have evolved into complex multi-scenario ecosystems where users engage through diverse behavioral contexts—browsing personalized feeds, conducting topical searches, discovering content creators, and responding to targeted advertisements. Each interaction scenario captures distinct yet complementary aspects of user intent: search queries reveal explicit informational needs, feed engagement demonstrates implicit content preferences, and advertisement clicks indicate commercial interests \citep{covington2016deep,hidasi2015session}. Crucially, users exhibit remarkably consistent underlying interests across these diverse behavioral contexts. A user passionate about sustainable living may search for "eco-friendly packaging," engage with environmental advocacy posts in their feed, and click advertisements for solar panels—each interaction revealing the same core interest through different behavioral lenses. This consistency suggests that user interests are inherently multi-dimensional, manifesting through intertwined behavioral trajectories that span multiple scenarios \citep{xia2020multiplex,zhu2022personalized}.

Despite this behavioral richness, production recommendation systems typically operate as independent silos, with separate models independently optimized for specific business objectives such as \ac{ctr} in feeds and \ac{advv} in advertisements \citep{zhang2019deep,chapelle2014simple}. This siloed design traps systems in local optima and creates several critical limitations. First, it fragments user understanding by restricting each model to narrow behavioral contexts, preventing holistic interest modeling. Second, it produces inconsistent user experiences when independent systems infer divergent preferences from the same user. Most importantly, it underutilizes valuable cross-scenario signals, limiting knowledge transfer across tasks and weakening performance for users with sparse activity in certain scenarios \citep{xia2020multiplex,zhu2022personalized}. Consider our sustainable living enthusiast: traditional systems would treat their search behavior, feed engagement, and ad responses as unrelated signals, failing to synthesize these coherent signals into unified user interest and intent.

\begin{figure}[t!]
    \centering
    \includegraphics[width=\linewidth]{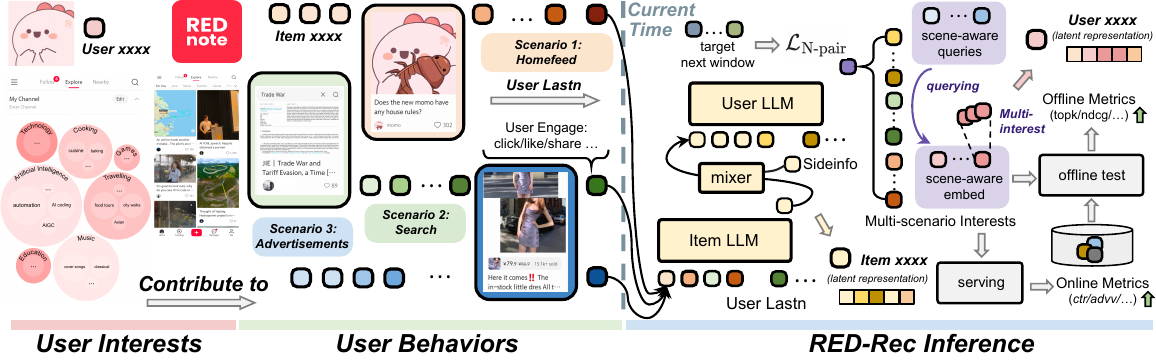}
    \caption{\textbf{From fragmented signals to unified understanding.} Users express consistent interests across diverse scenarios (left), generating rich behavioral sequences that span homefeed browsing, search queries, and ad interactions (middle). \acs{method} synthesizes these cross-scenario signals using \acs{llm}-powered hierarchical modeling to generate comprehensive user representations for context-aware recommendations (right). Vector graphics; zoom for details.}
    \label{fig:intro}
\end{figure}

We are motivated by the observation that users exhibit consistent interest patterns across diverse scenarios, and that modeling these patterns holistically can significantly enhance recommendation quality. While some cross-scenario modeling approaches exist \citep{zhang2023hierarchical,bao2023tallrec}, they typically require extensive manual feature engineering and struggle with scalability and robustness in production environments. Recent advances make this vision increasingly feasible: \acp{llm} have transformed semantic understanding of user behaviors and content \citep{wang2024mllm4rec}, while advanced sequence modeling techniques effectively capture complex temporal dynamics and cross-scenario dependencies \citep{sun2019bert4rec}. Meanwhile, modern \ac{ugc} platforms generate massive cross-scenario behavioral logs \citep{mcauley2015image,harper2015movielens,gao2022kuairand}, creating unprecedented opportunities for unified modeling at scale.

However, realizing this vision presents significant challenges: heterogeneity in action schemas, temporal dynamics, and semantics across scenarios; severe activity imbalances where users may have thousands of feed interactions but only dozens of searches; large-scale training and serving with strict latency and throughput constraints requiring sub-millisecond response times; and reconciling differing optimization objectives within a single architecture. While recent work explores mixtures of multi-source signals \citep{ma2022mixed,zhang2022multi,liu2024unified,yang2024mlora}, truly end-to-end unified modeling for industrial deployments remains underexplored.

We introduce \acf{method}, an \acs{llm}-enhanced hierarchical sequential recommendation framework tailored for billion-scale \ac{ugc} platforms. \ac{method} unifies interest modeling across heterogeneous contexts. First, we employ \ac{llm}-powered user and item encoders within a hierarchical two-tower structure that enables rich semantic representations while preserving efficiency for large-scale retrieval. Second, we introduce a novel 2D dense mixing policy that fuses cross-scenario behavioral signals along temporal and scenario axes to capture cross-scenario dependencies, coupled with multi-interest, scenario-aware queries that express fine-grained, context-specific user intents during serving. We train \ac{method} end-to-end on billions of behavioral events drawn from billions of items and over one hundred million users, incorporating system-level optimizations that enable near-real-time online deployment.

To enable rigorous evaluation, we also introduce a new cross-scenario sequential dataset curated from anonymized user behavior data on \platform, a world-leading UGC platform. The \ac{dataset} dataset spans millions of items and diverse user behaviors across feeds, search, and advertisement contexts, facilitating comprehensive benchmarking of unified and scenario-specific models. In a series of offline experiments, \ac{method} consistently outperforms strong baselines across multiple metrics and scenarios. This effectiveness extends to production, as demonstrated by online A/B testing, leading to a comprehensive full-scale rollout that now serves hundreds of millions of daily users on \platform.

Our main contributions include: (i) a unified, user-centric interest modeling framework that achieves both expressiveness and efficiency for billion-scale cross-scenario recommendation; (ii) a million-scale cross-scenario sequential dataset, \acs{dataset}, enabling rigorous evaluation of unified modeling approaches; and (iii) empirical validation in both offline and online production environments that establishes the practical viability of unified cross-scenario modeling at unprecedented scale.

\section{Related Work}

\paragraph{Sequential Recommendation}

Sequential recommendation has evolved from early neural methods like neural collaborative filtering \citep{he2017neural} and factorization machines \citep{rendle2010factorization,guo2017deepfm} to sophisticated sequence models. GRU4Rec \citep{hidasi2015session} pioneered recurrent architectures for session-based interactions, while Caser \citep{tang2018personalized} employed convolutional filters for temporal patterns. Transformer-based approaches like SASRec \citep{kang2018self} and BERT4Rec \citep{sun2019bert4rec} introduced self-attention and bidirectional encoding to capture long-range dependencies in user behavior sequences.
Recent advances address the multifaceted nature of user preferences through multi-interest modeling \citep{li2019multi,cen2020controllable}, graph neural networks \citep{wang2020global,zhang2022dynamic,yang2023dgrec}, and contrastive learning \citep{zhou2020s3,wei2023multi}. The emergence of \acp{llm} has opened new frontiers with enhanced user and item representations \citep{chen2024hllm,hu2024enhancing,wang2024mllm4rec} and generative paradigms \citep{chen2024enhancing,paischer2024preference,deng2025onerec,han2025mtgr}, though most remain limited to smaller-scale applications.

\paragraph{Cross-Scenario Modeling}

Users maintain consistent interests across different behavioral contexts despite varying interaction patterns \citep{zang2022survey}. Early cross-platform studies \citep{niu2021heterogeneous,tan2021multi} established that users exhibit similar topical preferences across platforms, motivating disentangled representation learning that separates stable interests from context-dependent behaviors.
Modern cross-scenario systems \citep{tan2021multi,zhao2023cross,li2024scene,chen2024survey,wu2025user} capture shared interest representations while accommodating scenario-specific patterns. Graph-based approaches \citep{tan2021multi,cao2022contrastive} model multi-behavioral patterns, while cross-domain \citep{ma2022mixed} and multi-domain methods \citep{zhao2023cross,yang2024mlora} leverage multi-source user histories to improve performance across scenarios.
However, most existing methods struggle with industrial-scale deployment challenges, including heterogeneity, scale, and strict latency requirements. Recent foundation model approaches \citep{wang2024llm4msr,shen2024exploring} show promise but lack validation at billion-scale scenarios. Our work addresses these limitations through an \acs{llm}-enhanced framework designed for industrial deployment with comprehensive online validation.

\section{The \acs{dataset} Dataset}

\begin{wrapfigure}{R}{0.55\linewidth}
    \centering
    \small
    \includegraphics[width=\linewidth]{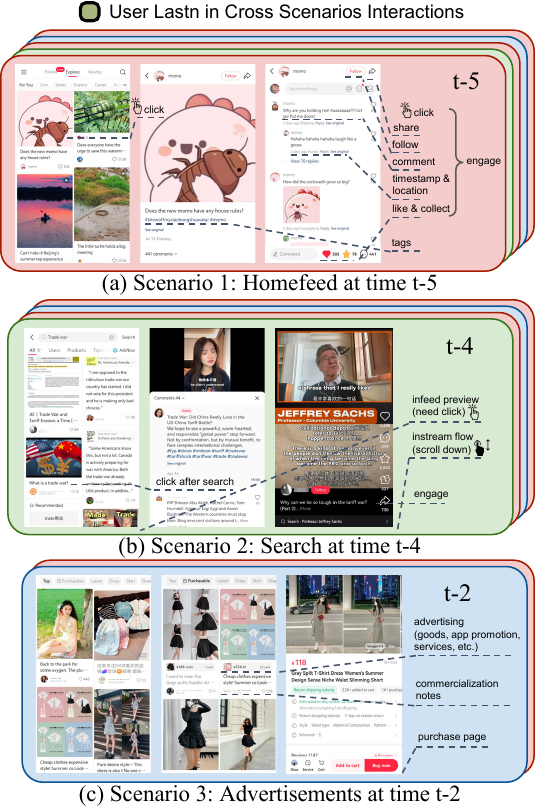}
    % \vspace{-18pt}
    \caption{\textbf{Cross-scenario user interactions in the \acs{dataset} dataset.} Real platform interface examples showing user behavior progression across (a) homefeed content consumption, (b) search-based exploration, and (c) advertisement engagement. Each scenario captures diverse interaction types, including clicks, shares, comments, and purchases, illustrating the interconnected nature of cross-scenario interactions on \ac{ugc} platforms.}
    \label{fig:data_example}
    \vspace{-60pt}
\end{wrapfigure}

Existing open-source sequential recommendation datasets suffer from major limitations in scope and diversity. Traditional datasets focus on isolated scenarios with singular interaction types such as ratings, clicks, or purchases \citep{ben2015recsys,harper2015movielens,ni2019justifying,zhu2018learning}, failing to capture the cross-scenario nature of modern \ac{ugc} platforms. Even recent datasets like KuaiRand \citep{gao2022kuairand} and Qilin \citep{chen2025qilin} that begin to characterize \ac{ugc} environments adopt fragmented approaches that underrepresent the complex interplay between scenarios and only partially reflect holistic user interest evolution.

To address these limitations, we introduce a cross-scenario sequential recommendation dataset, \acf{dataset}, derived from anonymized user data spanning billions of interactions on a major \ac{ugc} platform. Details on the protection of user privacy by excluding personal information, hashing user identifiers, and retaining only essential behavioral signals in the \ac{dataset} dataset are described in \cref{sec:supp:dataset:privacy}. Our dataset features the following three key characteristics:

\paragraph{Diverse Behavioral Contexts}

The \ac{dataset} dataset encompasses comprehensive real-world interaction scenarios, including homefeed browsing, search-driven exploration, and advertisement engagement. This temporally aligned diversity enables robust analysis of user behavior across distinct yet interconnected scenarios within a unified platform ecosystem.

\paragraph{Rich Engagement Patterns}

The \ac{dataset} dataset captures both explicit positive engagements (clicks, likes, collections, shares) and negative signals, along with view duration for each interaction. This provides a nuanced and holistic depiction of user preferences and attention patterns beyond simple binary feedback.

\paragraph{Industrial-Scale Coverage}

The dataset includes billions of items and over one hundred million users' engagement records, surpassing existing datasets in both scale and complexity. Tracking user behavior over extended time periods facilitates the study of long-context interest evolution, behavioral stability, and cross-scenario consistency that are typically unavailable in public datasets.

\begin{figure}[t!]
    \centering
    \includegraphics[width=\linewidth]{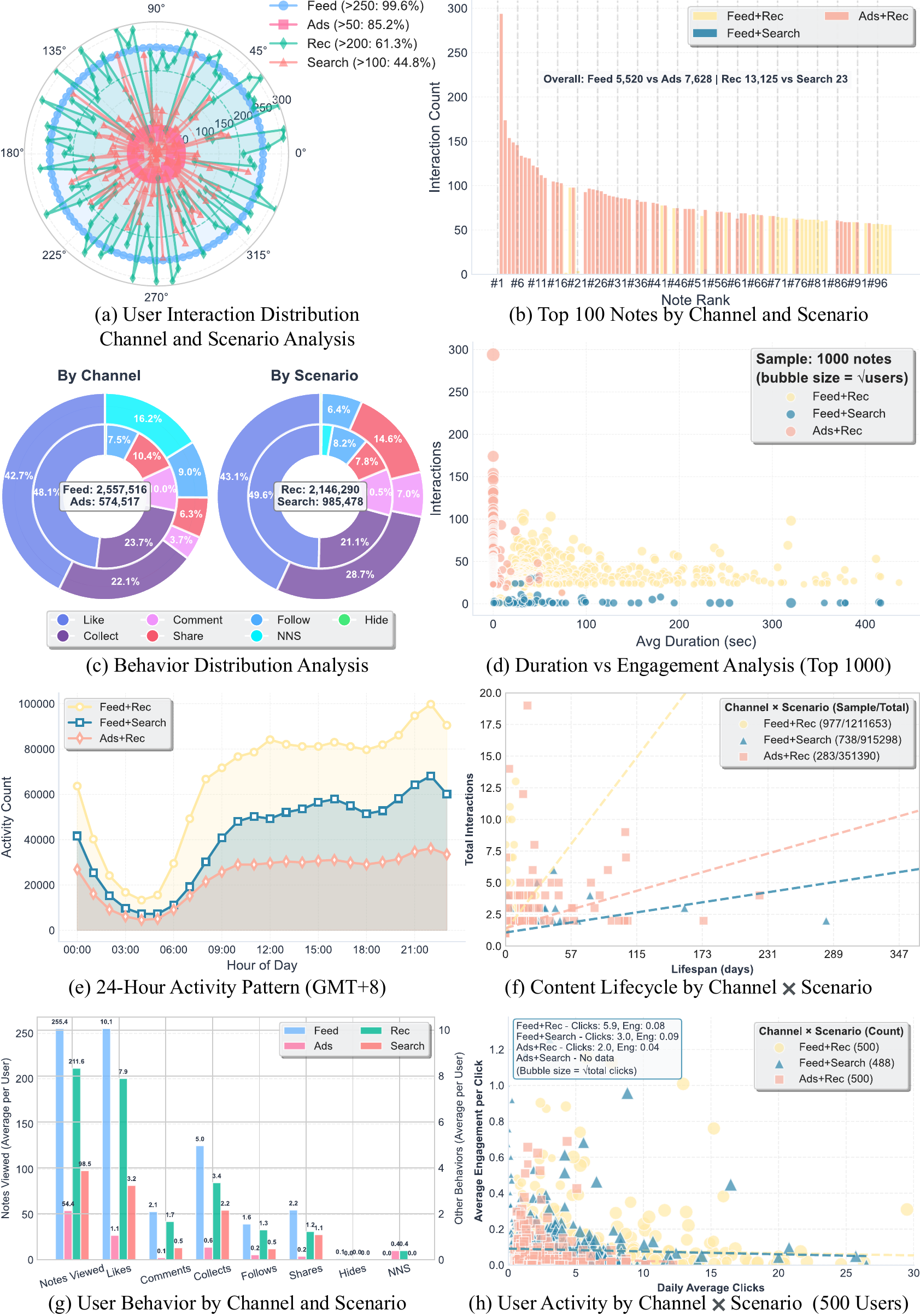}
    \caption{\textbf{Comprehensive user engagement analytics across scenarios.} Multi-faceted analysis of 10k sampled users showing interaction patterns across homefeed, advertisements, and search scenarios. The dashboard presents: (a) user distribution and interaction channels, (b) top content engagement by scenario, (c) behavioral pattern distributions, (d) duration-engagement correlations, (e) temporal activity patterns, (f) content lifecycle analysis, (g-h) cross-scenario user behavior comparisons. Analysis reveals distinct engagement patterns and temporal dynamics across different recommendation contexts.}
    \label{fig:dataset_statistic}
\end{figure}

\cref{fig:data_example} shows an example datapoint across multiple scenarios, including homefeed, search, and advertisements. \cref{fig:dataset_statistic} presents overall dataset statistics of \ac{dataset}, showing details on user engagement analytics. Additional details, including dataset collection and filtering, can be found in \cref{sec:supp:dataset}.

\section{Cross-Scenario User Interests Learning}\label{sec:model}

\subsection{Task Formulation}

We formulate cross-scenario sequential recommendation as learning unified user and item representations from cross-scenario interaction data. This formulation captures the complexity of modern recommendation systems where users engage with content across multiple behavioral contexts within the same platform ecosystem.

\paragraph{Problem Setup}

Consider a recommendation system with user set $\mathcal{U} = \{u_1, u_2, \ldots, u_N\}$ and universal item space $\mathcal{I} = \{i_1, i_2, \ldots, i_M\}$. The item space encompasses diverse content types, including image-text posts and videos created by regular users or advertisers, which can be recommended through homefeed, discovered via search, or presented as advertisements.

\paragraph{Cross-Scenario Interaction Sequences}

For each user $u \in \mathcal{U}$, we observe a chronologically ordered engagement sequence:
\begin{equation}
    S_u = \{(i_1, a_1, s_1, t_1), (i_2, a_2, s_2, t_2), \ldots, (i_{|S_u|}, a_{|S_u|}, s_{|S_u|}, t_{|S_u|})\}.
\end{equation}
Each interaction tuple contains: (i) $i_t \in \mathcal{I}$ - the interacted item, (ii) $a_t \in \mathcal{A}$ - the engagement action, (iii) $s_t \in \mathcal{S}$ - the scenario context, and (iv) $t$ - the interaction timestamp.

We focus on three primary scenarios $\mathcal{S} = \{$homefeed, advertisements, search$\}$ that represent the dominant user engagement patterns on \ac{ugc} platforms. User actions span a rich set of engagements $\mathcal{A} = \{$like, share, comment, follow, messaging, block$\}$, capturing both positive and negative feedback signals that reflect nuanced user preferences.

\paragraph{Learning Objective}

Our goal is to learn unified embedding functions that capture user preferences and item characteristics across different scenarios. Specifically, we aim to learn (i) a user embedding function: $f_u: \mathcal{U} \times \mathcal{H}_u \rightarrow \mathbb{R}^d$, and (ii) an item embedding function: $f_i: \mathcal{I} \rightarrow \mathbb{R}^d$. These functions map users (conditioned on their interaction history $\mathcal{H}_u$) and items to a shared $d$-dimensional embedding space:
\begin{equation}
    \mathbf{u} = f_u(u, S_u), \quad \mathbf{v}_i = f_i(I).
\end{equation}
The resulting embeddings $\mathbf{u}, \mathbf{v}_i \in \mathbb{R}^d$ encode cross-scenario user interests and item characteristics, enabling effective recommendation across all scenarios. These representations can be directly applied to recall tasks or serve as features for downstream ranking models.

\subsection{Hierarchical \acs{llm}-Based Representation Learning}

\ac{method} employs a hierarchical two-tower architecture that learns comprehensive user and item representations across multiple scenarios. The framework consists of three key components: (i) multimodal item encoding that captures content semantics, (ii) sequential user modeling with cross-scenario interest fusion, and (iii) scenario-aware querying mechanism for diverse preference capture.

\paragraph{Multimodal Item Representation}

Each item $i \in \mathcal{I}$ is encoded through a multimodal encoder $\text{E}_{\text{item}}$ that processes both textual and visual content:
\begin{equation}
    \mathbf{e}_i = \text{E}_{\text{item}}(\mathbf{x}_i, \mathbf{v}_i; \theta_t, \theta_v) \in \mathbb{R}^d.
\end{equation}
The textual component $\mathbf{x}_i$ encompasses title, tags, content description, and OCR-extracted text, processed by a pre-trained language model with parameters $\theta_t$. Visual content $\mathbf{v}_i$ is encoded using a ViT \citep{dosovitskiy2020image} with parameters $\theta_v$, followed by linear projection to dimension $d$. This unified representation captures rich semantic information across modalities.

\paragraph{Cross-Scenario Sequential Modeling}

\begin{figure}[t!]
    \centering
    \includegraphics[width=\linewidth]{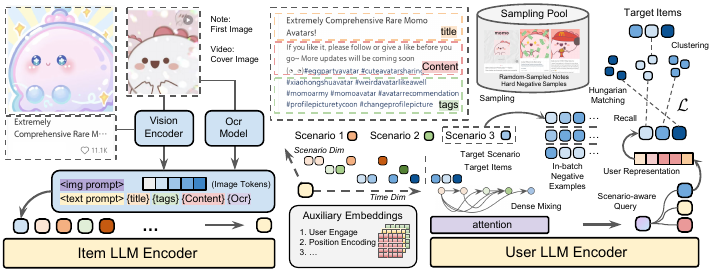}
    \caption{\textbf{Overall framework of \acs{method}.} \acs{method} employs a two-tower hierarchical architecture with multimodal item encoding and cross-scenario user modeling. The item encoder processes textual content (title, tags, OCR) and visual signals through unified embeddings. The user encoder incorporates a 2-D dense mixing policy to balance interactions across homefeed, advertisements, and search scenarios, followed by scenario-aware transformer blocks that capture evolving user interests. During training, positive and negative samples are drawn from a sampling pool to optimize contrastive objectives end-to-end. \acs{method} generates unified representations suitable for cross-scenario recommendation tasks.}
    \label{fig:model}
\end{figure}

To model user interests across diverse behavioral contexts, we aggregate interactions from three primary scenarios. For user $u$, the combined interaction sequence is $S_u = S_u^h \cup S_u^a \cup S_u^s$, where $S_u^h$, $S_u^a$, and $S_u^s$ represent homefeed, advertisements, and search respectively.
Each interaction incorporates three information dimensions: (i) \textbf{Content} represented by item embeddings $\mathbf{H}_u = [\mathbf{e}_{i_1}, \mathbf{e}_{i_2}, \ldots, \mathbf{e}_{i_n}] \in \mathbb{R}^{n \times d}$, (ii) \textbf{Actions} encoding engagement behaviors $\mathbf{A}_u = [\mathbf{a}_{i_1}, \mathbf{a}_{i_2}, \ldots, \mathbf{a}_{i_n}]$ as dense embeddings from one-hot vectors representing $\{\text{collect, share, message, block, like}\}$, and (iii) \textbf{Temporal} features with hour-level timestamps $\mathbf{h}_{i_t} = \text{OneHot}(\text{hour}(t)) \in \{0,1\}^{24}$ converted to dense embeddings. Hence, $\hat{\mathbf{H}}_u$ combines all three dimensions, enabling the user encoder to capture temporal patterns and engagement preferences.

\paragraph{2-D Dense Mixing Policy}

To address behavioral imbalance across scenarios, we introduce a balanced sampling strategy that preserves informative signals from all scenarios:
\begin{equation}
    S_u^{\text{mixer}} = \mathrm{Merge}\Big( S_u^{\text{homefeed}}[-n_h:],\ S_u^{\text{advertisements}}[-n_a:],\ S_u^{\text{search}}[-n_s:] \Big).
\end{equation}
The $\mathrm{Merge}(\cdot)$ operation chronologically sorts and concatenates recent interactions while maintaining scenario tags. This ``2-D dense mixing'' filters along both scenario (quota balancing) and temporal (recency) dimensions, ensuring representation of infrequent but valuable user signals. We design 2-D positional encoding for each event $j$ in $S_u^{\text{mixer}}$: $\mathbf{p}_{j} = \mathbf{PE}_{\text{seq}}(j) + \mathbf{PE}_{\text{gap}}(\Delta t_j)$, where $\mathbf{PE}_{\text{seq}}(j)$ captures sequence position and $\mathbf{PE}_{\text{gap}}(\Delta t_j)$ encodes time gaps with $\Delta t_j = t_{\text{curr}} - t_j$.

\paragraph{Scenario-Aware Interest Querying}

To capture diverse facets of user preferences, we employ learnable query embeddings $\mathbf{Q} = [\mathbf{q}_1, \mathbf{q}_2, \ldots, \mathbf{q}_K] \in \mathbb{R}^{K \times d}$ that attend to different interest aspects across scenarios. The scenario-aware representation is computed as:
\begin{equation}
    \mathbf{U}_u^{\text{query}} = \mathrm{E}_{\text{user}}\left(\left[\tilde{\mathbf{H}}_u[:-W];\, \mathbf{Q}\right];\, \theta_u\right),
\end{equation}
where $W$ represents the window size for recent interactions, enabling the model to generate multiple interest-specific representations.

\paragraph{Training Objective}

We optimize using Noise Contrastive Estimation (NCE) with temperature scaling:
\begin{equation}
    \mathcal{L}_{\mathrm{NCE}} = -\sum_{u,t} \log \frac{\exp\left(\tau \cdot \cos(\mathbf{u}_{u,t},\, \mathbf{v}_{i_{t+1}})\right)}{\exp\left(\tau \cdot \cos(\mathbf{u}_{u,t},\, \mathbf{v}_{i_{t+1}})\right) + \sum_{j \in \mathcal{N}} \exp\left(\tau \cdot \cos(\mathbf{u}_{u,t},\, \mathbf{v}_j)\right)},
\end{equation}
where $\tau$ is a learnable temperature parameter, $\mathbf{u}_{u,t}$ represents user $u$'s interest at time $t$, and $\mathcal{N}$ contains negative samples. Additionally, we incorporate window-based contrastive loss for recent interactions to capture evolving preferences. Similar techniques have also been employed in HyMiRec \citep{zhou2025hymirechybridmultiinterestlearning}, where hybrid multi-interest learning is applied for enhanced user representation and retrieval. Rather than a single, biased embedding, we encourage the User LLM model to capture diverse user intents from multiple perspectives.

\subsection{Evaluation Protocol}

We evaluate learned representations on recall tasks using temporal data splitting. For each user $u$, interactions are divided at randomly sampled cutoff $t_{\text{cut}}$, creating input sequence $S_u^{\text{input}} = \{(i, a, s, t) \in S_u : t < t_{\text{cut}}\}$ and target set $\mathcal{G}_u = \{ i_{t_{\text{cut}}},\ i_{t_{\text{cut}}+1},\ i_{t_{\text{cut}}+2} \}$.

The candidate pool $\mathcal{C}$ combines random platform samples with ground truth targets. User embeddings $\mathbf{u}$ computed from $S_u^{\text{input}}$ generate similarity scores $\text{score}(u, i) = \cos(\mathbf{u}, \mathbf{v}_i)$ for ranking. We report Hit Rate (HR@K), Normalized Discounted Cumulative Gain (NDCG@K), and Mean Reciprocal Rank (MRR) for $K \in \{10, 50, 100, 1000\}$.

\section{Experiments}\label{sec:exp}

\subsection{Experimental Setup}

\paragraph{Dataset and Configuration}

We conduct comprehensive experiments on an industrial dataset comprising 1 million users for training and 10,000 test samples for evaluation. The candidate pool contains approximately 1 million randomly sampled notes to ensure fair comparison across methods. Our default configuration sets window size $W=10$, sequence length $\text{last\_n}=128$, and employs 3 queries per scenario to capture diverse interest facets.

\paragraph{Model Architecture and Training}

Both item and user encoders are initialized with large language models: either 1.3B-parameter Chinese-LLaMA \citep{Chinese-LLaMA-Alpaca} or 1.5B Qwen-2.5 \citep{yang2025qwen3}, while visual encoding utilizes CLIP ViT-B/16. Training is conducted on 8 NVIDIA H100 GPUs for 3 epochs with batch size 2 and gradient accumulation of 4, requiring approximately 24 hours. Implementation details are provided in \cref{supp:sec:details}.

\begin{table}[t!]
    \centering
    \small
    \setlength{\tabcolsep}{3pt}
    \caption{\textbf{Single-scenario recommendation performance comparison.} Performance evaluation of \acs{method} variants against established baselines (SASRec, MoRec, HSTU, HLLM, DLRM-v3) on homefeed and advertisement recommendation tasks. \acs{method} variants include symbol-based (\ac{method}-symbol) and multimodal (\acs{method}-mm) versions, with pre-trained variants (\acs{method}-pt) leveraging large-scale data. Higher scores indicate better performance across all metrics.}
    \label{tab:results_single_scenario}% 
    \resizebox{\linewidth}{!}{%
        \begin{tabular}{lcccccccc}
            \toprule
            & \multicolumn{4}{c}{\textbf{Homefeed}} 
            & \multicolumn{4}{c}{\textbf{Advertisements}} \\
            \cmidrule(lr){1-2} \cmidrule(lr){2-5} \cmidrule(lr){6-9}
            \textbf{Baselines} \color{red} $\uparrow$ & HR/NDCG$_{10}$ & HR/NDCG$_{100}$ & HR/NDCG$_{1k}$ & MRR$_{*100}$
            & HR/NDCG$_{10}$ & HR/NDCG$_{100}$ & HR/NDCG$_{1k}$ & MRR$_{*100}$ \\
            SASRec & 1.76/0.97 & 12.32/1.79 & 32.01/4.04 & 1.01 & 3.26/1.63 & 14.08/3.71 & 39.11/5.27 & 1.57 \\ 
            MoRec & 1.78/\textbf{1.25} & 12.48/\textbf{2.23} & 31.98/\textbf{4.12} & 1.21 & 3.47/1.67 & 13.98/\textbf{3.88} & 38.27/4.89 & \textbf{1.78} \\
            HSTU   & \textbf{1.79}/1.22 & 12.72/2.21 & 31.76/3.69 & 1.15 & 3.85/1.70 & 14.32/3.30 & 38.20/\textbf{5.38} & 1.43 \\ 
            HLLM & 1.66/0.62 & \textbf{12.77}/1.83 & \textbf{32.52}/4.02 & \textbf{1.22} & \textbf{4.21/1.21} & 14.27/3.37 & \textbf{39.21}/4.48 & 1.39 \\ 
            DLRM-v3 & 1.63/1.03 & 11.33/2.01 & 28.96/3.72 & 1.13 & 3.54/1.21 & \textbf{15.27}/3.22 & 35.39/4.27 & 1.67 \\
            \midrule
            \ac{method} & 2.31/0.68 & 12.59/1.88 & 31.94/3.86 & 1.27 & 4.24/1.28 & 16.44/3.21 & 40.18/4.61 & 1.96 \\
            \acs{method}-pt & 2.90/0.63 & 14.89/2.02 & 36.16/4.01 & 1.30 & \textbf{4.84/1.30} & 17.66/2.87 & \textbf{42.71/5.21} & \textbf{2.27} \\
            \ac{method}-mm & 2.35/1.21 & 14.20/\textbf{2.27} & 31.29/3.97 & 1.29 & 4.31/1.31 & 17.22/3.18 & 41.86/4.66 & 1.92\\
            \ac{method}-mm-pt & \textbf{3.23/1.27} & \textbf{15.46}/2.21 & \textbf{36.29/4.14} & \textbf{1.38} & 4.82/1.19 & \textbf{18.21/3.29} & 42.56/4.98& 2.21 \\
            \bottomrule
        \end{tabular}%
    }%
\end{table}

\paragraph{Baselines and Evaluation}

We compare against established recommendation methods: SASRec \citep{kang2018self}, MoRec \citep{yuan2023go}, HSTU \citep{zhai2024actions}, HLLM \citep{chen2024hllm}, and DLRM-v3 \citep{naumov2019deep}. Our evaluation encompasses both single-scenario (homefeed, advertisements) and cross-scenario (search + homefeed, homefeed + advertisements, all combined) settings. We assess four \ac{method} variants: \ac{method}-symbol, \ac{method}-mm, and their pre-trained versions (\ac{method}-symbol-pt, \ac{method}-mm-pt) trained on large-scale online data. Standard metrics (HR@K, NDCG@K, MRR) are reported across multiple cutoff values.

\subsection{Single-Scenario Performance}

\begin{wraptable}{r}{0.55\linewidth}
    \vspace{-30pt}
    \centering
    \small
    \setlength{\tabcolsep}{3pt}
    \caption{\textbf{Cross-scenario recommendation performance evaluation.} Performance comparison when leveraging cross-scenario signals for improved recommendations across different target scenarios. Results demonstrate the effectiveness of \ac{method} in utilizing cross-scenario behavioral signal. Higher scores indicate superior performance.}
    \label{tab:results_vertical_scenarios}
    \resizebox{\linewidth}{!}{%
        \begin{tabular}{lcccc}%
            \toprule
            \multicolumn{5}{c}{\textbf{Search + Homefeed} (for Homefeed)} \\
            \midrule
            \textbf{Baselines} \color{red} $\uparrow$ & HR/NDCG$_{10}$ & HR/NDCG$_{100}$ & HR/NDCG$_{1k}$ & MRR$_{*100}$ \\
            SASRec        & 1.73/1.22 & 12.02/3.21 & 32.17/4.17 & 1.52 \\
            MoRec         & 1.79/1.30 & 13.92/2.99 & 33.01/3.98 & 1.53 \\
            HSTU          & 1.79/1.25 & 12.84/3.28 & 33.15/4.24 & 1.55 \\
            HLLM          & 1.69/1.02 & 13.49/3.18 & 33.04/4.21 & 1.58 \\
            DLRM-v3       & 1.64/1.18 & 11.35/3.02 & 30.89/3.98 & 1.48 \\
            \acs{method}   & 2.26/1.32 & 14.74/3.16 & 33.29/4.20 & 1.58 \\
            \acs{method}-pt& \textbf{2.92/1.33} & \textbf{18.26/3.24} & \textbf{38.92/4.23} & \textbf{1.67} \\
            \midrule[1pt]
            \multicolumn{5}{c}{\textbf{Homefeed + Advertisements} (for Advertisements)} \\
            \midrule
            \textbf{Baselines} \color{red} $\uparrow$ & HR/NDCG$_{10}$ & HR/NDCG$_{100}$ & HR/NDCG$_{1k}$ & MRR$_{*100}$ \\
            SASRec        & 3.72/1.24 & 16.18/3.08 & 38.94/4.72 & 1.94 \\
            MoRec         & 3.80/1.30 & 17.23/2.62 & 38.29/4.77 & 1.98 \\
            HSTU          & 3.89/1.28 & 16.95/3.15 & 40.12/4.81 & 2.01 \\
            HLLM          & 3.68/1.19 & 17.24/3.12 & 39.76/4.78 & 1.97 \\
            DLRM-v3       & 3.52/1.21 & 15.43/2.95 & 36.87/4.58 & 1.87 \\
            \acs{method}   & 4.36/1.31 & 18.32/3.27 & 42.61/5.02 & 2.11 \\
            \acs{method}-pt& \textbf{5.18/1.38} & \textbf{18.89/3.21} & \textbf{46.59/5.57} & \textbf{2.38} \\
            \midrule[1pt]
            \multicolumn{5}{c}{\textbf{Homefeed + Search + Advertisements} (for Advertisements)} \\
            \midrule
            \textbf{Baselines} \color{red} $\uparrow$ & HR/NDCG$_{10}$ & HR/NDCG$_{100}$ & HR/NDCG$_{1k}$ & MRR$_{*100}$ \\
            SASRec        & 3.68/1.21 & 14.29/2.08 & 38.94/4.72 & 1.94 \\
            MoRec         & 3.82/1.33 & 18.27/2.98 & 38.41/4.66 & 1.98 \\
            HSTU          & 3.92/1.31 & 17.21/3.19 & 40.14/4.81 & 2.11 \\
            HLLM          & 4.08/1.11 & 19.92/3.18 & 43.27/4.91 & 2.06 \\
            DLRM-v3       & 3.34/1.01 & 14.08/2.81 & 35.74/4.36 & 1.74 \\
            \acs{method}   & 4.72/1.33 & 18.33/3.22 & 42.89/4.97 & 1.94 \\
            \acs{method}-pt & \textbf{5.18/1.35} & \textbf{20.52/3.24} & \textbf{49.17/5.93} & \textbf{2.41} \\
            \bottomrule
        \end{tabular}%
    }%
    \vspace{-18pt}
\end{wraptable}

\cref{tab:results_single_scenario} presents results for homefeed and advertisement recommendation scenarios. \ac{method} consistently outperforms all baselines in both scenarios, demonstrating the efficacy even in single-scenario settings. The superior performance stems from two key factors: (i) multi-interest user representation learning that captures diverse preference facets, and (ii) advanced \acs{llm}-based semantic encoding that provides richer representations than traditional ID-based methods.

Compared to SASRec and HSTU, which rely on item ID embeddings, \ac{method} shows substantial improvements, particularly beneficial for cold-start scenarios where semantic understanding is crucial. When compared against HLLM, which shares similar architectural principles, \ac{method} benefits from larger backbone models with enhanced Chinese language capabilities, enabling better alignment with our dataset characteristics and further performance gains.

\subsection{Cross-Scenario Benefits}

Cross-scenario evaluation (\cref{tab:results_vertical_scenarios}) reveals significant performance improvements when integrating information across behavioral contexts. The most pronounced gains occur in two key scenarios: (i) search data enhancing homefeed recommendations, and (ii) combined homefeed and search signals improving advertisement performance.

\paragraph{Cross-Scenario Information Flow}

Incorporating cross-scenario signals consistently improves performance across all baselines, with \ac{method} achieving the largest gains due to its effective integration capabilities and advanced user-side \ac{llm} reasoning. For homefeed recommendations, access to search behaviors, particularly post-search engagement patterns, substantially increases both HR and NDCG scores. Similarly, advertisement recommendations benefit from combined homefeed and search behaviors, showing the greatest metric improvements across all evaluation criteria.

\paragraph{Consistent Improvements}

\begin{figure}[t!]
    \centering
    \begin{subfigure}[t]{0.48\linewidth}
        \centering
        \includegraphics[width=\linewidth]{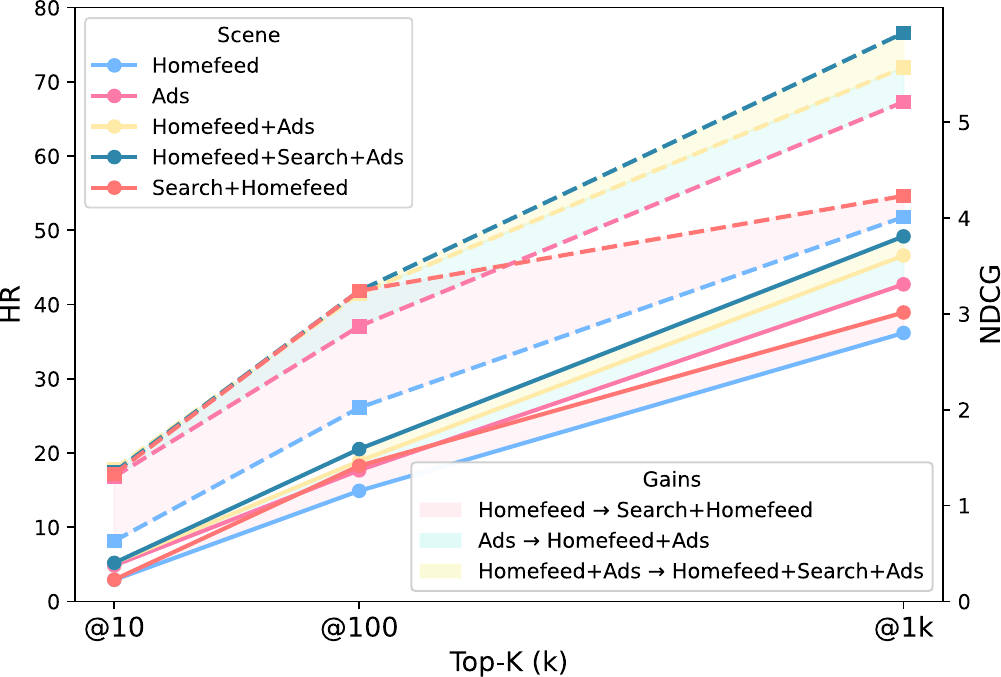}
        \caption{HR and NDCG for different scenarios (with gains highlighted).}
        \label{fig:combined_a}
    \end{subfigure}%
    \hfill%
    \begin{subfigure}[t]{0.48\linewidth}
        \centering
        \includegraphics[width=\linewidth]{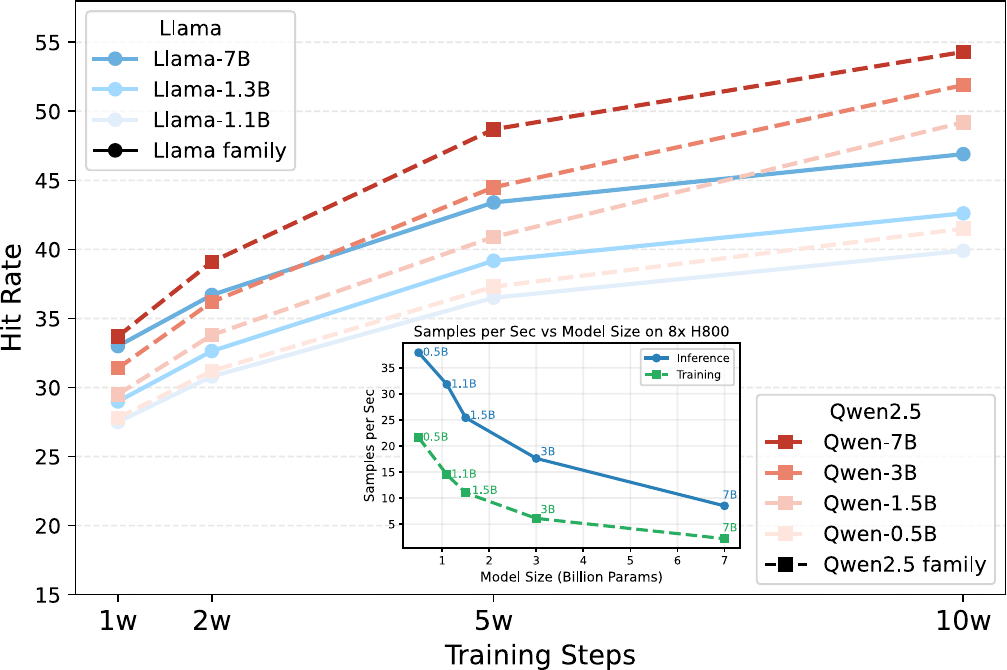}
        \caption{Scaling law: HR \vs training steps for LLaMA and Qwen models.}
        \label{fig:combined_b}
    \end{subfigure}%
    \caption{\textbf{Cross-scenario benefits and model scaling analysis.} (a) Performance improvements from cross-scenario modeling using \ac{method}, where different colored lines represent various scenario combinations (homefeed, ads, search) and the shaded regions highlight the performance gains achieved through unified cross-scenario learning compared to single-scenario baselines. (b) Scaling laws for model size versus Hit Rate performance, showing consistent improvements with increased parameter count across both LLaMA and Qwen model families, with corresponding serving throughput (samples per second) trade-offs indicated by the secondary axis.}
    \label{fig:combined}
\end{figure}

These enhancements remain consistent across all cutoff values ($K \in \{10, 50, 100, 1000\}$), indicating that \ac{method} not only increases the likelihood of relevant item recommendation but also improves their ranking positions. \cref{fig:combined} illustrates these cross-scenario benefits, demonstrating substantial performance gains when leveraging complementary signals.

\begin{wraptable}{r}{0.65\linewidth}
    \vspace{-27pt}
    \centering
    \small
    \setlength{\tabcolsep}{3pt}
    \caption{\textbf{Ablation study results for \ac{method} architectural components.} \textbf{Top:} Core model configuration ablations examining the impact of sequence length (SeqLen = 128 vs. 32) and interest modeling approaches (Multi Interest vs. Single Interest) on recommendation performance. \textbf{Bottom:} Cross-scenario mixing policy ablations evaluating different strategies for combining behavioral signals across homefeed, search, and advertisement scenarios. Comparison includes temporal sampling, scenario exclusion variants, and our proposed 2D Dense Mixing approach. Results demonstrate the effectiveness of longer sequences, multi-interest modeling, and comprehensive cross-scenario signal integration.}
    \label{tab:ablation}
    \resizebox{\linewidth}{!}{%
        \begin{tabular}{lcccc}%
            \toprule
            \multicolumn{5}{c}{\textbf{Homefeed}} \\
            \midrule
            \textbf{Setting} & HR/NDCG$_{10}$ & HR/NDCG$_{100}$ & HR/NDCG$_{1k}$ & MRR \\
            \midrule
            SeqLen = 128, Multi-Interest, pt & \textbf{2.90}/0.63 & \textbf{14.89}/\textbf{2.02} & \textbf{36.16}/4.01 & 1.30 \\
            SeqLen = 128, Multi-Interest & 2.31/0.68 & 12.59/1.88 & 31.94/3.86 & 1.27 \\
            SeqLen = 128, Single-Interest & 1.85/\textbf{0.72} & 10.24/1.95 & 26.78/\textbf{4.12} & \textbf{1.31} \\
            SeqLen = 64, Multi-Interest & 2.08/0.71 & 11.32/1.94 & 28.67/3.92 & 1.29 \\
            SeqLen = 32, Multi-Interest & 1.72/0.61 & 9.48/1.76 & 25.47/3.64 & 1.29 \\
            \midrule[1pt]
            \multicolumn{5}{c}{\textbf{Homefeed + Search + Advertisements}} \\
            \midrule
            \textbf{Mixer Strategy} & HR/NDCG$_{10}$ & HR/NDCG$_{100}$ & HR/NDCG$_{1k}$ & MRR \\
            \midrule
            Sorted by Timestamp        & 2.10/0.53    & 10.55/1.90   & 21.44/2.20   & 0.65 \\
            Naive Combination          & 4.28/1.22    & 17.60/3.06   & 41.90/4.85   & 1.85 \\
            1D (on position)  & 4.31/1.23    & 17.65/3.08   & 41.95/4.87   & 1.86 \\
            1D (on timestamp)  & 4.40/1.25    & 17.80/3.10   & 42.20/4.90   & 1.88  \\
            2D-Mixing (\ac{method})             & \textbf{4.72/1.33} & \textbf{18.33/3.22} & \textbf{42.89/4.97} & \textbf{1.94} \\
            \bottomrule
        \end{tabular}%
    }%
    \vspace{-15pt}%
\end{wraptable}

\subsection{Ablation Studies}

We conduct comprehensive ablation studies examining key architectural components and design choices (\cref{tab:ablation}). Our analysis focuses on four critical aspects: (i) input sequence length effects, (ii) multi-interest query mechanisms, (iii) large-scale pretraining benefits, and (iv) cross-scenario mixing strategies.

\paragraph{Core Component Analysis}

Results demonstrate that longer input sequences, multi-interest queries, and large-scale pretraining all contribute to improved recommendation metrics. The multi-interest querying mechanism proves particularly valuable, enabling the model to capture diverse facets of user preferences across different scenarios. Large-scale pretraining provides substantial performance gains, highlighting the importance of leveraging extensive behavioral data.

\paragraph{Cross-Scenario Mixing Strategies}

For cross-scenario settings, our 2D dense mixing policy achieves the strongest performance compared to alternative fusion methods. This validates our approach of integrating both positional and temporal information for effective signal combination, addressing behavioral imbalance while preserving informative signals from all scenarios.

\subsection{Scaling Analysis}

To balance model accuracy with deployment efficiency, we investigate scaling laws across different model sizes. We train models from both LLaMA \citep{touvron2023llama} (0.5B-7B) and Qwen \citep{yang2025qwen3} (0.5B-7B) families on identical token volumes and evaluate on our test set.

\cref{fig:combined_b} presents Hit Rate performance and corresponding serving throughput (\ac{sps}) for the Homefeed+Search+Advertisements scenario. Results show consistent HR improvements with increased model size up to 7B parameters across both families, indicating potential scaling benefits. However, larger models significantly reduce serving throughput, creating practical deployment constraints. The 1.5B Qwen-2.5 model represents an optimal balance between performance and efficiency for production deployment. 

\subsection{Online Deployment}

We validate \ac{method} through online A/B testing in the recall stage of \platform’s advertising recommendation system. The experiment uses balanced traffic allocation (10\% treatment vs. 10\% control) over one week, evaluating performance against the entire item catalog comprising items distributed within the past two months (about 1.1 billion items). Recall results from our method are incorporated as an additional recall channel. The deployment operates in near real time: for each user, we gather and truncate their most recent N interactions, perform inference to generate a user-side embedding, and retrieve relevant items by matching it with precomputed item-side embeddings, thus enabling timely and effective recall.

\ac{method} achieves significant improvements: 0.8864\% increase in total \ac{advv} and 0.3401\% boost in overall \ac{cost}. These gains are particularly noteworthy given the platform's scale. Notably, over 90\% of the items selected during the initial candidate generation phase are uniquely contributed by this recall path, demonstrating that our approach provides significant incremental recommendations. The significant gains in advertising scenarios further validate our offline findings regarding cross-scenario knowledge transfer from homefeed patterns. Encouraged by these promising results, we deployed \ac{method} platform-wide, now serving approximately 160 million daily users. This deployment demonstrates the practical viability of \acs{llm}-based cross-scenario recommendation at industrial scale, offering considerable business value while maintaining acceptable serving performance.

\section{Conclusion}

We present \ac{method}, a unified hierarchical \acs{llm}-based framework that addresses cross-scenario sequential recommendation at an industrial scale. Our key contributions include: (i) a two-tower architecture integrating multi-modal content understanding with cross-scenario behavioral modeling, (ii) scenario-aware mixing and multi-interest querying mechanisms that capture diverse user preferences, and (iii) comprehensive validation demonstrating substantial gains in both offline experiments and production deployment. Our comprehensive empirical evaluations, encompassing experiments on both offline multi-scenario dataset and large-scale real-world deployment, consistently demonstrate substantial gains over strong baselines in both offline and production settings. Our work demonstrates that unified user interest modeling across behavioral contexts is both technically feasible at scale and essential for coherent, user-centric recommendations. By bridging diverse interaction patterns, \ac{method} enables more seamless personalized content discovery and offers new insights for cross-domain recommendation research and industrial applications.

\subsubsection*{Ethics Statement}

The data utilized in our model training and the constructed dataset have been fully anonymized to protect user privacy. The dataset contains only interactions with publicly accessible content and excludes all personally identifiable information. All data collection and processing procedures adhere to relevant privacy regulations and platform policies. We acknowledge that recommendation systems can potentially introduce algorithmic bias and filter bubbles, and encourage practitioners to implement appropriate fairness monitoring and mitigation strategies when deploying such systems. 

\bibliography{reference_header,references}
\bibliographystyle{iclr2026_conference}
\clearpage

\appendix
\renewcommand\thefigure{A\arabic{figure}}
\setcounter{figure}{0}
\renewcommand\thetable{A\arabic{table}}
\setcounter{table}{0}
\renewcommand\theequation{A\arabic{equation}}
\setcounter{equation}{0}
\pagenumbering{arabic}% resets `page` counter to 1
\renewcommand*{\thepage}{A\arabic{page}}
\setcounter{footnote}{0}

\section{Terminology}\label{sec:supp:term}

We would like to first offer additional explanations for specific terminology used throughout the paper in order to facilitate understanding for non-expert readers:

\paragraph{Homefeed} refers to the main feed or landing page displayed to a user when they open a content platform or app. It typically consists of a personalized selection of items (such as posts, products, videos, etc.) recommended to the user based on their preferences and past behavior.

\paragraph{Internal Flow} denotes the content consumption pattern within the single-column sliding or swiping through content (\eg, images, videos, or articles). Users engage with recommendations directly within this detailed view by navigating between related items or sliding to the next recommended content.

\paragraph{External Flow} refers to the content consumption flow that occurs on the main feed of the platform, where users browse the list of recommended items presented to them upon opening the app. This process typically involves users scrolling vertically through the two-column page.

\paragraph{Scenarios} refer to distinct user interaction environments or channels within the platform, each characterized by unique user intents and behavioral patterns. In this paper, we focus on three core scenarios: \textbf{homefeed}, \textbf{advertisements}, and \textbf{search}. The homefeed scenario represents the primary personalized feed where users consume a diverse assortment of recommended content. The advertisement scenario corresponds to user engagement with sponsored or promotional content distributed throughout various parts of the platform. Although advertisement content can appear within the homefeed, we treat it as a separate scenario because it represents a different source and serves distinct business objectives. The search scenario involves users actively retrieving information or content by submitting queries.  

\paragraph{last-$n$} refers to the most recent `n' items a user has interacted with on the platform. For example, `last10' indicates the user's last 10 consumed items. This concept is commonly used to capture and analyze a user's most current interests or activity history.

\paragraph{Engage} represents user interactions with content, such as clicks, likes, comments, shares, or dwell time. Engagement metrics are used to measure how users interact with recommended items and to assess the effectiveness of recommender systems.

\section{Further experiments}\label{sec:supp:furexp}

We further test \ac{method} on Amazon Books Reviews \citep{mcauley2015image}, a widely used subset in recommender system research datasets, which was sampled from the Amazon Review dataset. In the Books subset, each review typically contains fields such as reviewer ID, item (book) ID, rating (1-5 stars), review text, timestamp, and sometimes additional metadata (\eg, book title). We test and compare \ac{method} in \cref{tab:fur_exp}.

\begin{table}[ht!]
    \centering
    \small
    \caption{\textbf{A comparison of the Amazon Books dataset.}}
    \begin{tabular}{cccc}
        \toprule
        \textbf{Baselines} \color{red} $\uparrow$ & \textbf{HR/NDCG$_{10}$} & \textbf{HR/NDCG$_{50}$} & \textbf{HR/NDCG$_{200}$} \\
        \midrule
        SASRec    & 3.06/1.64 & 7.54/2.60 & 14.31/3.62  \\
        MoRec(bert)         &    3.21/1.82       &  8.21/2.33  &  18.29/3.71        \\
        HSTU-1B      & 4.78/2.62 & 10.82/3.93 & 19.08/5.17  \\
        DLRM-v3-1B   & 6.22/2.88 & 12.74/5.12 & 23.12/5.29 \\
        HLLM-1B      & 9.28/5.65 & 17.34/7.41 & 27.22/\textbf{8.89}  \\
        \ac{method}-1.1B-LLaMA    & 9.46/5.49 & 18.63/7.03 & 29.88/8.45  \\
        \ac{method}-1.5B-Qwen2.5    & \textbf{9.98/5.88} & \textbf{19.98/7.88} & \textbf{32.62}/8.78  \\
        \bottomrule
    \end{tabular}
    \label{tab:fur_exp}
\end{table}

\section{\ac{dataset} Dataset}\label{sec:supp:dataset}

\subsection{Dataset Statistics}

We compare our training dataset with other existing datasests or benchmarks from \ac{ugc} platforms in \cref{supp:table:data_comp}.

\begin{table}[ht!]
    \centering
    \small
    \caption{\textbf{A brief comparison of public-released datasets and the training dataset \acs{method} used.}}
    \begin{tabular}{cccccc}
        \toprule
        \textbf{Property} & \textbf{Amazon} & \textbf{JD Search} & \textbf{KuaiSAR} & \textbf{Qilin} & \textbf{\ac{method}(training)}\\
        \midrule
        Users        & 192.4k    & 173.8k     & 25.8k     & 15.5k   & 1.0m   \\
        Items        & 63.0k     & 12.9m  & 6.9m  & 2.0m   &  300.6m \\
        Queries      & 3.2k      & 171.7k     & 453.7k    & 571.9k  & search items  \\
        Actions      & 1.7m  & 26.7m  & 19.7m & 2.5m  & 683.2m  \\
        Content   & text/image & text & text/video & text/image/video & text/image/video   \\
        Scenario & Rec & Search & Search+Rec & Search+Rec & Search+Rec+Ads \\
        \bottomrule
    \end{tabular}
    \label{supp:table:data_comp}
\end{table}

An item in the training data, and also in the proposed \ac{dataset} dataset is like:

\begin{lstlisting}[caption={Example of an item in the training dataaset.},frame=single]
{
  "user_id": "xxxx",
  "data": {
    "homefeed_item_lastn": [
      {
        "duration": 28,
        "is_click": 1,
        "is_click_profile": 0,
        "is_collect": 0,
        "is_comment": 0,
        "is_follow": 0,
        "is_hide": 0,
        "is_like": 0,
        "is_nns": 0,
        "is_pagetime": 1,
        "is_read_comment": 1,
        "is_share": 0,
        "is_videoend": 0,
        "item_id": "684a48440000000023014319",
        "page_key": 0,
        "timestamp": 1749771247,
        "type": "note"
      },
      {
        "duration": 17,
        "is_click": 1,
        "is_click_profile": 0,
        "is_collect": 0,
        ...
        "is_pagetime": 1,
        "is_read_comment": 0,
        "is_share": 0,
        "is_videoend": 0,
        "item_id": "684aa0720000000021003dbe",
        "page_key": 0,
        "timestamp": 1749732355,
        "type": "note"
      }
      // ...
    ],
    "ads_item_lastn": [ ... ],
    "search_item_lastn": [ ... ]
  }
}
\end{lstlisting}
where
\begin{itemize}
    \item \textbf{user\_id}: Unique identifier for the user, \eg, \texttt{xxxx}.
    \item \textbf{data}:
    \begin{itemize}
        \item \textbf{homefeed\_item\_lastn}: An array of objects representing the last $n$ items from the user's home feed. Each object contains:
        \begin{itemize}
            \item \texttt{duration}: Viewing duration (in seconds).
            \item \texttt{is\_click}: Whether the item was clicked ($1$) or not ($0$).
            \item \texttt{is\_click\_profile}: Whether the user's profile was clicked ($1$ or $0$).
            \item \texttt{is\_collect}: Whether the item was collected or saved ($1$ or $0$).
            \item \texttt{is\_comment}: Whether the item was commented on ($1$ or $0$).
            \item \texttt{is\_follow}: Whether the user followed from this item ($1$ or $0$).
            \item \texttt{is\_hide}: Whether the item was hidden ($1$ or $0$).
            \item \texttt{is\_like}: Whether the item was liked ($1$ or $0$).
            \item \texttt{is\_message}: Whether the author of the message was messaged ($1$ or $0$).
            \item \texttt{is\_pagetime}: Whether the page time event was triggered ($1$ or $0$).
            \item \texttt{is\_read\_comment}: Whether comments were read ($1$ or $0$).
            \item \texttt{is\_share}: Whether the item was shared ($1$ or $0$).
            \item \texttt{is\_videoend}: Whether a video was watched until the end ($1$ or $0$).
            \item \texttt{item\_id}: Identifier for the content item.
            \item \texttt{page\_key}: Page identifier.
            \item \texttt{timestamp}: Timestamp of the interaction.
            \item \texttt{type}: Type of item, \eg, \texttt{note}.
        \end{itemize}
        \item \textbf{ads\_item\_lastn}: Array of the last $n$ interacted advertisement items (\texttt{item\_id}, \texttt{duration}, etc.).
        \item \textbf{search\_item\_lastn}: Array of the last $n$ search items with similar structure.
    \end{itemize}
\end{itemize}

\subsection{Privacy and Validation}\label{sec:supp:dataset:privacy}

User privacy is strictly protected in our dataset by excluding all personal or sensitive user information beyond anonymized behavior sequences. User identifiers are securely hashed to prevent any possibility of re-identification, and all content items featured in the dataset are publicly available, with no private materials included. Furthermore, only essential behavioral signals required for recommendation research are retained, while potentially identifying metadata such as device information and location is omitted. Engagement timestamps are also consistently biased to prevent reconstruction of individual timelines. Together, these measures ensure the dataset enables recommendation research without compromising user confidentiality or privacy.

We focus exclusively on active platform users who demonstrate substantial engagement patterns: users must have at least 30 valid clicks in the homefeed scenario and 5 valid clicks in the advertisement scenario, where a click is considered valid only if the associated viewing duration exceeds 5 seconds.

\section{Metrics}\label{sec:supp:metrics}

In this work, we focus on three widely adopted metrics: Hit Ratio (HR), Normalized Discounted Cumulative Gain (NDCG), and Mean Reciprocal Rank (MRR). In recommender systems and information retrieval, model performance is typically assessed by ranking-based evaluation metrics that reflect both the accuracy and the ordering of recommendations. These metrics are evaluated at various ranking cutoffs $K$ (\eg, $K=10, 100, 1000$) to provide a comprehensive view of retrieval quality across different user engagement depths.

\paragraph{Hit Ratio (HR)}

Hit Ratio (HR@K) measures the proportion of test cases in which at least one relevant item, usually the ground-truth item, is found within the top-$K$ positions of the ranked recommendation list. Formally, for a set of $N$ users (or queries), it is defined as:
\begin{equation}
    \mathrm{HR}@K = \frac{1}{N} \sum_{i=1}^N \mathbb{I}(\mathrm{rank}_{i} \leq K),
\end{equation}
where $\mathrm{rank}_i$ denotes the position (starting from 1) at which the ground-truth item for the $i$-th user occurs in the predicted ranking, and $\mathbb{I}(\cdot)$ is the indicator function. HR is equivalent to recall@K in the case of a single relevant item per query.

HR@K is intuitive and interpretable, indicating the likelihood that a user's desired item appears among the top-$K$ recommendations. However, it does not reward higher placements within the top-$K$ and disregards the relative ranking among recommended items.

\paragraph{Normalized Discounted Cumulative Gain (NDCG)}

Normalized Discounted Cumulative Gain (NDCG@K) extends HR@K by accounting for the position of relevant items, rewarding items that are ranked higher in the recommended list. For each test case, DCG is computed as:
\begin{equation}
    \mathrm{DCG}@K = \sum_{j=1}^{K} \frac{\mathrm{rel}_{ij}}{\log_2(j+1)},
\end{equation}
where $\mathrm{rel}_{ij}$ is the relevance label (typically 1 for the ground-truth item and 0 otherwise) for the $j$-th item in the ranked list for user $i$. The DCG is then normalized by the ideal DCG (IDCG), i.e., the maximum possible DCG for that user, to yield:
\begin{equation}
    \mathrm{NDCG}@K = \frac{1}{N} \sum_{i=1}^N \frac{\mathrm{DCG}_i@K}{\mathrm{IDCG}_i@K}
\end{equation} 

NDCG@K captures both the relevance and ranking quality, penalizing relevant items that appear lower in the ranking. It is especially useful in scenarios with multiple relevant items per user or graded relevance.

\paragraph{Mean Reciprocal Rank (MRR)}

Mean Reciprocal Rank (MRR@K) evaluates how highly the first relevant item is ranked, and is defined as:
\begin{equation}
    \mathrm{MRR}@K = \frac{1}{N} \sum_{i=1}^{N}\frac{1}{\mathrm{rank}_i},
\end{equation}
where $\mathrm{rank}_i$ is the position of the first relevant item in the recommended list for user $i$, and set to infinity (i.e., reciprocal rank is 0) if no relevant items are found in the top-$K$. MRR@K emphasizes early precision, heavily rewarding algorithms that surface the relevant item at or near the top. Its sensitivity to the first relevant item's position makes it particularly apt for settings prioritizing immediate relevance (\eg, question answering, search).

\paragraph{Evaluation Protocols and Cutoff Values}

In our work, all metrics above are computed at different cutoff values $K$ to approximate various user scenarios (\eg, users interacting with the top 10 or top 100 items). These are denoted as $\mathrm{HR}@K$, $\mathrm{NDCG}@K$, and $\mathrm{MRR}@K$, for various $K$ (\eg, $K=10,100,1000$). For interpretability and easier comparison, MRR is often multiplied by 100 and reported as $\mathrm{MRR}_{*100}$. These metrics are computed under a leave-one-out or leave-many-out evaluation: for each user, one or more ground-truth relevant items are held out (used as positives), and the ranking is judged over a candidate pool comprising these positives and many sampled negatives.

\section{Implementation Details}\label{supp:sec:details}

We provide additional implementation details of the proposed \ac{method}.

\subsection{Item Encoder}

The item encoder is designed to construct robust content representations, leveraging a pretrained \ac{llm} as its foundation. Textual information related to each item---including titles and descriptions---is concatenated, tokenized, and prepended with a designated special token to sharpen the representation focus. This sequence is then passed through the \ac{llm} encoder, producing dense semantic embeddings for each item. Specifically, we extract the embedding corresponding to the special token. The resulting embedding's dimension matches the model's hidden size; for instance, 1536 for LLaMA2-1.3B and 3584 for Qwen-7B.

For multimodal input, there are essentially two primary approaches. The first involves utilizing an individual vision encoder, such as ViT\citep{dosovitskiy2020image}, like LLaVA\citep{liu2023visual}, to extract visual tokens, which are then projected into the language embedding space. The second approach directly leverages vision-language models (VLMs) such as Qwen-VL, which jointly process visual and textual inputs within a unified architecture. In our work, we primarily adopt the first approach based on considerations of model size and efficiency for online serving.

\subsection{User Encoder}

User representation learning is managed via hierarchical interest modeling over long interaction histories. User interaction sequences are first encoded using the item encoder, resulting in contextualized item embeddings. These are then organized and refined by the proposed mixer module that captures temporal and sequential dependencies. The enhanced representations are subsequently fed into a disentangled multi-interest learning module, which extends beyond conventional single-vector user profiles by learning multiple independent embeddings, each attending to a distinct facet of user intent.

Training supervision extends past traditional next-item prediction, encompassing all interactions within a lookahead window to better reflect realistic browsing patterns. To achieve this, we apply cosine similarity clustering to partition target items based on behavioral signals, followed by the Hungarian algorithm matching to associate each cluster centroid with its corresponding interest vector. A contrastive loss function drives the specialization of each embedding, ensuring broad coverage and effective disambiguation of diverse user preferences across multiple interest groups. Complete implementation details are available in our supplementary code repository.

To model user interests in a disentangled manner, we introduce learnable queries that capture refined, distinct interests according to three key principles: sufficient supervision for each query, minimal overlap in interest coverage, and coherent optimization directions. Given refined interest embeddings $\{\mathbf{r}_1,\ldots,\mathbf{r}_s\}$ and positive samples $\{\mathbf{t}_1,\ldots,\mathbf{t}_w\}$ from the target window, we cluster the positive samples into $s$ groups using cosine similarity and then match cluster centroids to interest embeddings via the Hungarian algorithm to maximize pairwise similarity. The contrastive loss is applied only to these matched pairs:
\begin{equation}
    \mathcal{L}_{\text{total}} = \frac{1}{w} \sum_{i=1}^{w} \sum_{j=1}^{s} \mathcal{L}_{\text{ctr}}(t_i, r_j) \cdot \Pi(i, j),
\end{equation}
where $\Pi(i, j)=1$ if the cluster of $t_i$ is matched with $r_j$, and $0$ otherwise. The contrastive loss $\mathcal{L}_{\text{NCE}}$ is defined as:
\begin{equation}
    \mathcal{L}_{\text{NCE}}(t, r) = -\log \frac{e^{\text{sim}(t, r)/\tau}}{e^{\text{sim}(t, r)/\tau} + \sum_{i=1}^{m} e^{\text{sim}(r, e_i)/\tau}},
\end{equation}
where $m$ is the number of negative samples, $e_i$ is the $i$th negative sample embedding, and $\text{sim}$ denotes cosine similarity.

This design enables adaptive learning: queries naturally specialize for users with diverse interests and converge for users whose preferences are more focused.

\section{Additional Experiments}

\subsection{Pretraining Validation}

Our first set of experiments investigates the effect of varying the backbone \acp{llm} for the item and user encoders. Specifically, we explore the following configurations: (i) using different pretrained \acp{llm} for item and user encoders, (ii) training one or both encoders from scratch instead of initializing from a pretrained model, and (iii) freezing the item encoder during training. The detailed results are summarized in \cref{tab:llm_ablation}.

Across all settings, we observe that using exactly the same pretrained \ac{llm} for both item and user encoders and fine-tuning them jointly yields the best performance. In contrast, utilizing mismatched encoders, initializing from scratch, or freezing either encoder all result in significant drops in overall accuracy. This suggests that consistent representation spaces and co-adaptation between the two encoders are crucial for optimal model performance.

\begin{table}[t!]
    \centering
    \small
    \caption{\textbf{Ablation results for different combinations of item and user \acsp{llm} and training strategies.}}
    \label{tab:llm_ablation}
    \resizebox{\linewidth}{!}{
    \begin{tabular}{clcccc}
        \toprule
        \textbf{Scenario} & \textbf{Configuration} & HR/NDCG$_{10}$ & HR/NDCG$_{100}$ & HR/NDCG$_{1k}$ & MRR$_{*100}$ \\
        \midrule
        Homefeed & \acs{method} (2 * Qwen)  & 2.31/0.68 & 12.59/1.88 & 31.94/3.86 & 1.27  \\ 
         & Item \acs{llm} from scratch           & 0.00/0.00 & 0.00/0.00 & 0.03/0.01 & 0.00 \\ 
         & User \acs{llm} from scratch           & 0.00/0.00 & 0.03/0.01 & 1.32/0.21 & 0.01 \\ 
         & Item \acs{llm} frozen & 1.27/0.36 & 5.51/0.77 & 11.37/1.02 & 0.37  \\ 
         & User \acs{llm} frozen & 1.78/0.44 & 10.47/1.02 & 23.06/1.48 & 1.01   \\
         \midrule
         Homefeed + Ads & \acs{method} (2 * Qwen)      & 4.36/1.31 & 18.32/3.27 & 42.61/5.02 & 2.11 \\ 
         & Item \acs{llm} from scratch & 0.00/0.00 & 0.00/0.00 & 0.08/0.04 & 0.01       \\ 
         & User \acs{llm} from scratch & 0.00/0.00 & 0.00/0.00 &      1.01/0.07 & 0.03 \\ 
         & Item \acs{llm} frozen & 1.49/0.41 & 9.49/1.31 & 19.29/1.52 & 0.76 \\ 
         & User \acs{llm} frozen & 2.57/1.01 & 13.72/1.98 & 29.72/1.88 & 3.28 \\
         \midrule
         Homefeed + Ads & \acs{method} (2 * Qwen)      & 4.36/1.31 & 18.32/3.27 & 42.61/5.02 & 2.11 \\ 
          & \acs{method}-CoT (2 * Qwen)      & 4.46/1.35 & 18.78/3.60 & 44.61/5.01 & 2.15 \\ 
        \bottomrule
    \end{tabular}
    }
\end{table}

\subsection{CoT Validation}

We explore explainable recommendations based on \acs{cot}-based \citep{wei2022chain} explanations for the input layer in a cross-scenario setting. In this experiment, we introduce a Chain-of-Thought (CoT) auxiliary loss: beyond learning discriminative user and item encoders, we encourage explainable cross-scenario reasoning by forcing the user model to generate natural language rationales for each action: 
\begin{equation}
    \mathcal{L}_{\text{CoT}} = - \sum_{u \in \mathcal{U}} \sum_{t=1}^{|S_u|} \sum_{\ell=1}^{L_t} \log p_\phi(r_{t,\ell} \mid r_{t<\ell}, \mathbf{z}_{u,t}),
\end{equation}
where $p_\phi()$ denotes the probability, computed by a learnable language model head parameterized by $\phi$, of generating the $\ell$-th token $r_{t,\ell}$ of the rationale conditioned on the previous tokens $r_{t<\ell}$ and the contextualized user embedding $\mathbf{z}_{u,t}$ at interaction $t$. The overall training loss is then $\mathcal{L}_{\text{total}} = \mathcal{L}_{\text{NCE}} + \lambda_{\text{CoT}}\mathcal{L}_{\text{CoT}}$.

We use GPT 4.1 to generate CoT explanations. An example of the generated CoT explanation is like:

\textit{The user browsed multiple articles related to Switzerland on the homepage, such as ``Do you dare to guess how many days of sunshine in Switzerland?'' and ``What to wear for a trip to Switzerland next week?'' This indicates a clear interest in Switzerland. While previously recommended advertisements included those related to travel, they were not specifically targeted at Switzerland.}

\textit{Therefore, we recommend to the user the targeted ad ``Personal tested and useful! The ultimate transportation ticket map tool for traveling in Switzerland!'', as well as other advertisements related to traveling in Switzerland, such as ``Countdown to opening! The four legendary theme parks of Fiesch First Mountain'' and ``Interlaken sledding premium tips | Save 400 RMB instantly.''}

The CoT explanation module is particularly well-suited to the cross-scenario recommendation setting. By generating step-by-step rationales that account for user behaviors across different scenarios or domains, the model can provide contextually accurate and human-understandable justifications for its recommendations. This improves both transparency and user trust, crucial for scenario-aware systems. However, we observe that applying the CoT-based approach to large-scale datasets introduces significant challenges. The requirement to generate context-dependent rationales for every user interaction leads to substantially increased computational and memory costs. Given these limitations, we restrict our experiments to small-scale testing. The detailed results are summarized in \cref{tab:llm_ablation}. Including CoT data in training has led to certain improvements, but it does not outperform the pretrained model.

\end{document}

%% file: config.tex
\usepackage{graphicx} \graphicspath{{figures/}}
\usepackage{amsmath,amssymb,mathabx,mathtools,amsthm,nicefrac}
\usepackage[pagebackref,breaklinks,colorlinks]{hyperref}
\usepackage[capitalise,noabbrev,nameinlink]{cleveref}
\usepackage[skip=3pt,font=small]{subcaption}
\usepackage[skip=3pt,font=small]{caption}
\usepackage{acronym}
\usepackage{wrapfig}
\usepackage[dvipsnames,svgnames,x11names,table]{xcolor}
\usepackage{enumerate}
\usepackage{xspace}
\usepackage[misc]{ifsym}
\usepackage{titletoc}
\usepackage{verbatim,fancyvrb,listings}
\usepackage{booktabs,tabularx,colortbl,multirow,multicol,array,makecell,tabularray}

\makeatletter
\DeclareRobustCommand\onedot{\futurelet\@let@token\@onedot}
\def\@onedot{\ifx\@let@token.\else.\null\fi\xspace}
\def\eg{\emph{e.g}\onedot}

\def\vs{\emph{vs}\onedot}

\makeatother

\acrodef{ugc}[UGC]{User-Generated Content}
\acrodef{llm}[LLM]{Large Language Model}
\acrodef{method}[\textit{RED-Rec}]{\underline{\textit{R}}ecommender \underline{\textit{E}}ngine for \underline{\textit{D}}iversified scenarios}
\acrodef{cot}[CoT]{Chain-of-Thought}
\acrodef{sps}[SPS]{Sample per Second}
\acrodef{advv}[\textit{ADVV}]{Advertiser Value}
\acrodef{cost}[\textit{Cost}]{Feed Ad Spend}
\acrodef{ctr}[CTR]{Click-Through Rate}
\makeatletter
\@ifundefined{iclrfinalcopy}{%
    \acrodef{dataset}[\textsc{Red-MMU}]{Multi-Scenario Multimodal User Behaviors}
}{%
    \acrodef{dataset}[\textsc{Red-MMU}]{RedNote's Multi-Scenario Multimodal User Behaviors}
}
\makeatother
\makeatletter
\@ifundefined{iclrfinalcopy}{%
    \newcommand{\platform}{a world-leading UGC platform}
}{%
    \newcommand{\platform}{Xiaohongshu}
}
\makeatother